\documentclass[aps, prd, amsmath, floats, floatfix, twocolumn, superscriptaddress, nofootinbib, showpacs]{revtex4}

\usepackage{graphicx,color}
\usepackage{latexsym}
\usepackage{amsmath,amssymb}        
\usepackage[draft=false]{hyperref}
\usepackage{mathrsfs}
\usepackage{comment}
\usepackage{soul}
\usepackage{mathrsfs}
\definecolor{orange}{rgb}{1,0.5,0}
\DeclareSymbolFontAlphabet{\mathrsfs}{rsfs}
\DeclareMathAlphabet{\mathcal}{OMS}{cmsy}{m}{n}

\usepackage{ulem}

\begin{document}
\title{Cosmology on a Gravitational Wave Background}

\author{Tonatiuh Matos}\email{tonatiuh.matos@cinvestav.mx}
\affiliation{Departamento de F\'isica, Centro de Investigaci\'on y de Estudios Avanzados del IPN, A.P. 14-740, 07000 M\'exico D.F.,
  M\'exico.}

  \author{Luis A. Escamilla}
\email{luis.escamilla@icf.unam.mx}
\affiliation{Instituto de Ciencias F\'isicas, Universidad Nacional Aut\'onoma de M\'exico, Apdo. Postal 48-3, 62251 Cuernavaca, Morelos, M\'exico.}

  \author{Maribel Hernández}
\email{maribel.hernandez@nucleares.unam.mx}
\affiliation{Instituto de Ciencias Nucleares, Universidad Nacional Aut\'onoma de M\'exico, Apdo. Postal  70-543, Deleg. Coyoac\'an, C.P. 04510, CDMX, M\'exico.}

  \author{J. Alberto V\'azquez}
\email{javazquez@icf.unam.mx}
\affiliation{Instituto de Ciencias F\'isicas, Universidad Nacional Aut\'onoma de M\'exico, Apdo. Postal 48-3, 62251 Cuernavaca, Morelos, M\'exico.}

\begin{abstract}
It is a fact that the universe lives on a Gravitational Wave Background (GWB), which
it may be in the form of extra energy, which is not contained in Einstein's field equations. 
In \cite{Matos:2021jef}, a new model was developed to explain the current accelerating expansion of the 
universe where a GWB was incorporated by extending Einstein's equations to 
$R_{\mu\nu}-\frac{1}{2}Rg_{\mu\nu}+\frac{2\pi^2}{{\lambda}^2}g_{\mu\nu}=\kappa ^2 T_{\mu\nu}$, where ${\lambda}$ is 
the Compton wavelength of the graviton. 
In the present work we show that this extended form agrees very well with the observations of Cosmic chronometers, 
Baryon Acoustic Oscillations and Pantheon SN type Ia, reproducing the observational data with a $\Delta\chi^2=3.26$ 
in favor of the present model compared to the $\Lambda$CDM. The values favored by these observations 
are $\Omega_{\rm m} = 0.31 \pm 0.02$, $H_0=68 \pm 0.02$ Km/s/Mpc, $\Omega_{\rm k} = 0.001\pm 0.011$; 
we also find an excellent consistence of this model with the Cosmic Microwave Background and the Matter Power Spectrum.
We conclude that this model is an excellent alternative to explain the accelerating expansion of the universe 
without incorporating the cosmological constant.
\end{abstract}

\maketitle

\section{Introduction}

In the realm of cosmology, one of the most significant revelations of the past century was the observation 
that the universe is not only expanding but also experiencing accelerated expansion. This extraordinary 
finding defied our expectations and sparked research endeavors to comprehend the underlying causes driving 
this peculiar behavior. It is within this context that the concept of Dark Energy emerged as a compelling 
explanation for the accelerated expansion. However, the fundamental nature of Dark Energy still 
remains a perplexing 
enigma, and unraveling its mysteries continues to be a captivating pursuit. Despite the multitude of proposals 
and ideas aimed to decipher this phenomenon, we have not yet arrived to a fully convincing solution 
(see, for instance, \cite{Poulin:2023lkg, Bamba:2022jyz, Frusciante:2019xia}).

In a previous study \cite{Matos:2021jef}, a novel model dubbed as the Compton Mass Dark Energy (CMaDE)  
was introduced, whose main goal is to incorporate the quantum nature of the gravitational field
into Einstein's equations, which could also be considered as the Gravitational Waves Background (GWB),  
and it could be a viable  explanation of the accelerated expansion of the universe. 
Very recently it has been demonstrated by several 
observatories that the universe is immersed in a GWB \cite{NANOGrav:2023gor, Reardon:2023gzh, Xu:2023wog, Antoniadis:2023lym},
in this case the frequencies observed are of the order of nanohertz and their origin is still unknown.  
However, there is no clear justification for restricting the GWB solely to nanohertz frequencies and 
therefore we will consider their wavelength may be extended to other scales, specifically 
those on cosmological scales. In this context, we let the specific origin 
of the GWB for other works, but it is important to clarify that it is an additional energy not 
incorporated into the Einstein's equations a priori, which is intrinsically connected to the space-time metric. 
Thus, similar to the aforementioned study, the proposal is to incorporate the gravitational wave energy 
of spacetime into Einstein's equations. Gravitational waves and the mediator of the gravitational 
interaction, here for simplicity named as graviton, has zero mass. To study the universe we 
focus on frequencies on cosmological scales. In \cite{Matos:2021jef} (see 
also \cite{Escobar-Aguilar:2023ekv}) this energy was introduced into the Einstein equation to obtain
\begin{equation}\label{eq:Einstein}
R^{\mu\nu}-\frac{1}{2}g^{\mu\nu}R+\frac{2\pi^2}{{\lambda}^2} g^{\mu\nu}=\kappa^2 T^{\mu\nu},
\end{equation}
where $\kappa^2=8\pi G/c^4$; $G$ is Newton's gravitational constant and ${\lambda}$ is the 
cosmological Compton wavelength of the graviton, which for cosmological scales depends only on the time $t$ 
coordinate due to the expansion of the universe. For a similar approach, but with a 
completely different philosophy, see for example \cite{Li:2004rb,Cai:2007us}.

In a previous study \cite{Matos:2022jzf}, it was demonstrated that these equations not only provide 
an explanation for the accelerated expansion of the universe but may also exhibit an agreement 
with key cosmological observations such as the Mass Power Spectrum (MPS) and the Cosmic Microwave 
Background (CMB) radiation. However, it is worth noting that the aforementioned works 
\cite{Matos:2021jef} and \cite{Matos:2022jzf},  were performed by an approximation where the covariant derivative 
of the energy-momentum tensor ${T^{\mu\nu}}_{;\nu}$ vanished. Although this approximation 
results in a minimal violation of the general principle of covariance, it is important to address 
this issue. Therefore, in the present study, we aim to remove the approximation and evaluate the 
performance of the CMaDE model against the standard $\Lambda$CDM model by confronting it with background data.

\section{The CMaDE model}

\begin{figure*}
\centering
\includegraphics[width=0.35\textwidth]{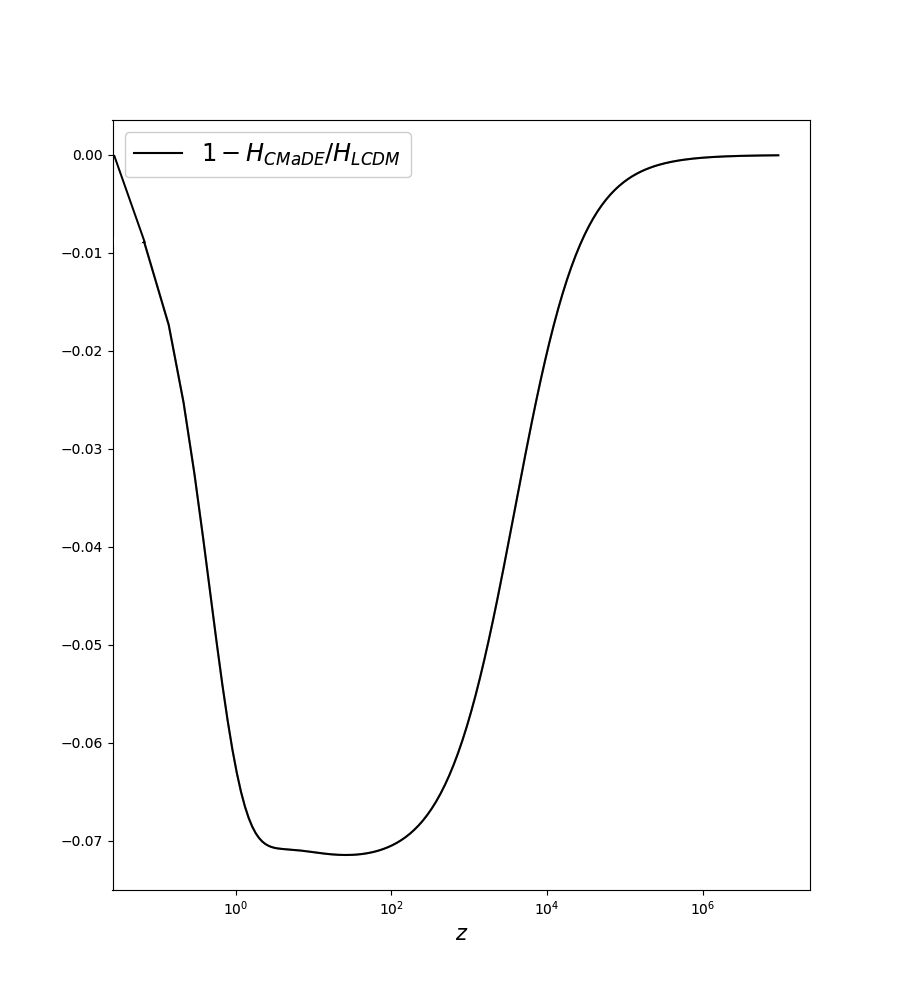}
\includegraphics[width=0.35\textwidth]{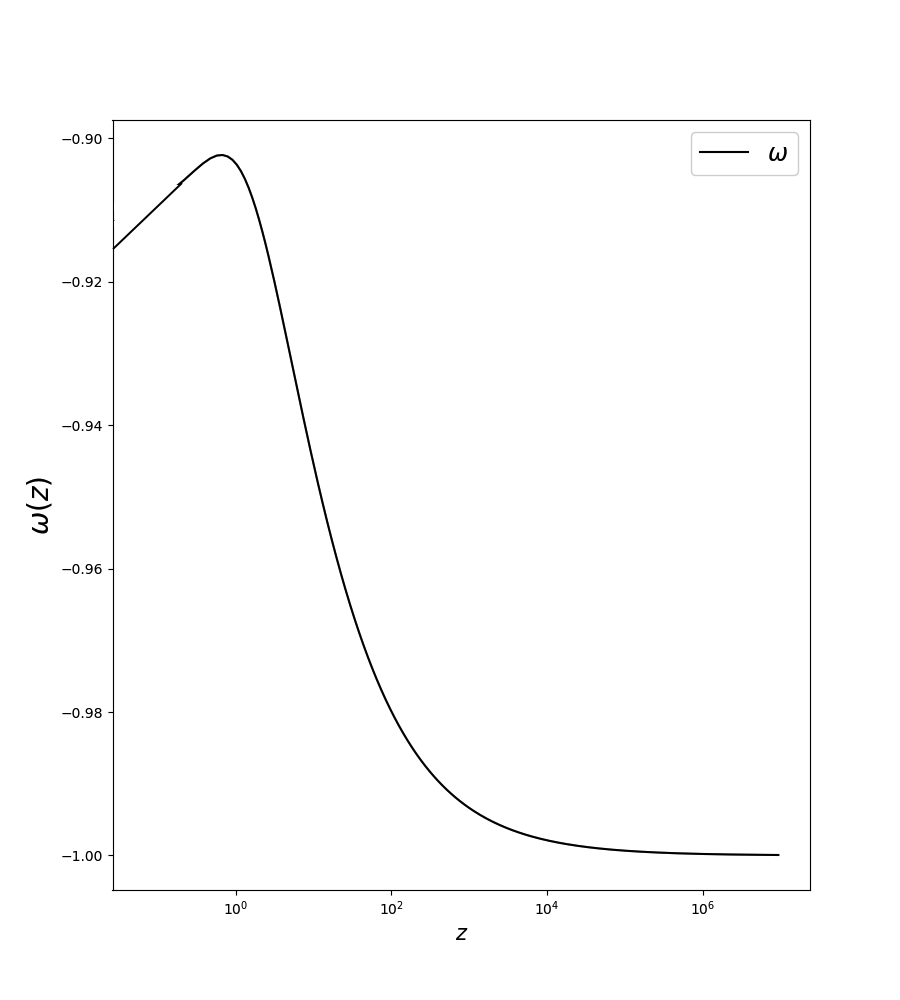}
\caption{In the left panel we show the relationship of the Hubble parameters through $1-H_{\rm{CMaDE}}/H_{\rm{LCDM}}$, 
from the solutions of the set of equations (\ref{eq:Sys}). We notice a similar behaviour, with a few percent 
level difference at small redshift. Here we set $\mathcal{H}_0=1$, $\Omega^0_{\rm{0b}}=0.044$, 
$\Omega^0_{\rm{0dm}}=0.27$, $\Omega^0_{\rm{0r}}=9.539\times10^{ -5} $ and $\Omega^0_{\rm{0k}}=0.08$ for both models. 
In the left panel we see the evolution of EoS (\ref{eq:omega}) in equation (\ref{eq:Sys}). Notice that the effective 
EoS converges to $-1$ at high redshifts, specially on the recombination epoch.}
\label{fig:Claudio}
\end{figure*}

In order to obtain the conservation equations of the system, in a FLRW universe, 
we perform the covariant derivative of equation (\ref{eq:Einstein}).  
Observe that ${G^{\mu\nu}}_{;\nu}=0$ by construction and ${g^{\mu\nu}}_{;\nu}=0$ because the 
metric is compatible with respect to the connection.
For index $\mu\not=0$ the covariant derivative of equation (\ref{eq:Einstein}) is an identity, 
but for $\mu=0$, we obtain
\begin{equation}
\label{eq:conservacion}
  \dot{\mathcal{M}} c^2=-\kappa^2\left(\dot\rho+3H(\rho+p)\right),
\end{equation}
where
\begin{equation}\label{eq:M}
    {\mathcal{M}}=\frac{2\pi^2}{\lambda^2},
\end{equation}
is the extra term in the Einstein's equations; where $H=\dot a/a$ is the Hubble parameter and $\rho$ 
and $p$ are the total energy density and pressure of the system, respectively. Here, we consider the 
components of the universe are the matter $\rho_{\rm m}$, which is made up of dark matter 
$\rho_{\rm dm}$ and baryons 
$\rho_{\rm b}$, and radiation $\rho_{\rm r}$ made up of neutrinos $\rho_{\nu}$ and photons $\rho_\gamma$. 
Bearing in mind we know the equations of state for baryons and radiation, hence 
$\dot\rho_{\rm b} +3H\rho_{\rm b}=0$ 
and $\dot\rho_{\rm r}+4H\rho_{\rm r}=0$, then we can plug in these results into equation (\ref{eq:conservacion}).
On the other hand, because the lack of information about the nature of the dark matter, as a first 
approximation we can assume it is made up of dust with $p_{\rm dm}=0$, just like baryons. The main 
difference with the previous works \cite{Matos:2021jef,Matos:2022jzf} is that here we let the dark matter 
interact with the GWB energy of space-time, avoiding the violation of the general covariance principle.
Furthermore, we know that $\rho=\rho_{\rm m}+\rho_{\rm r}=\rho_{\rm dm}+\rho_{\rm b}+\rho_{\rm r}$, 
and $p=p_{\rm r}+ p_{\rm m}=\frac{1}{3}\rho_{\rm r}$. 
Using these results, the equation (\ref{eq:conservacion}) becomes
\begin{eqnarray}\label{eq:rhos}
    k_c\dot{\mathcal{M}} c^2
    &=&-\kappa^2\left(\dot\rho_{\rm dm}+3H\rho_{\rm dm}\right),\nonumber\\
    \dot\rho_{\rm b}&=&-3H\rho_{\rm b},\nonumber\\
    \dot\rho_{\rm r}&=&-4H\rho_{\rm r},
\end{eqnarray}
where $k_c$ is a bias parameter that mediates the contribution of the space-time energy of 
the GWB to the dark matter.

Next we derive an equation for ${\mathcal{M}}$. We know that the wavelength ${\lambda}$ 
satisfies the relation ${\lambda}=(c/H_0)R_H$ \cite{Matos:2021jef}, with
\begin{equation}\label{eq:RH}
    R_H=H_0\int \frac{dt}{a}=H_0\int \frac{e^{-N}}{H}dN.
\end{equation}
It is convenient to use the e-folding coordinate $N$ defined as $N=\ln(a)$. We denote the derivative 
with respect to $t$ with an over dot, and a prime means the derivative with respect to $N$.
Then, from (\ref{eq:M}) we have that
\begin{eqnarray}
    {\mathcal{M}}'=-\frac{4\pi^2}{{\lambda}^3}{\lambda}'.
\end{eqnarray}
Thus, we use equation (\ref{eq:RH}) to obtain
\begin{equation}\label{eq:LambdaP}
    {\mathcal{M}}'=\pm\frac{\sqrt{2}c}{\pi}{\mathcal{M}}^{3/2}\frac{e^{-N}}{H}.
\end{equation}
However, different frequencies of the fluctuations may contribute in a different way  
to the accelerated expansion, that is, part of these fluctuations can be transformed into black holes, 
structures of the universe, etc. To mediate this contribution, we can set a bias parameter $Q$ 
in front of the equation (\ref{eq:LambdaP}).
Equations (\ref{eq:rhos}) together with the Friedmann equation
\begin{equation}\label{eq:Friedmann}
    H^2+\frac{k}{a^2}-\frac{{\mathcal{M}} c^2}{3}=\frac{\kappa^2\rho}{3},
\end{equation}
where $k=1,-1,0$, is the curvature parameter, are a complete set of equations for the 
variables $\rho_{\rm b}$, $\rho_{\rm r}$, $\rho_{\rm dm}$ and ${\mathcal{M}}$. It is convenient to rewrite these 
equations in terms of dimensionless quantities, then we introduce 
\begin{equation}
    \Omega_X^0=\frac{\rho_X\kappa^2}{3H_0^2}=\Omega^0_{0X}a^{-3(1+\omega_X)},
\end{equation}
for each component of the system and $\mathcal{H}=H/H_0$, where the second identity 
is valid only for barotropic fluids with $\omega_X=$ constant. Observe that in the 
definition of $\Omega_X^0$ we use $H_0$ instead of the traditional $H$. Therefore, 
we obtain a complete system of equations 
\begin{eqnarray}\label{eq:Sys}
\mathcal{H}^2&=&\Omega^0_{\rm b}+\Omega^0_{\rm dm}+\Omega^0_{\rm r}+\Omega^0_{0k}e^{-2N}+\Omega^0_{\mathcal{M}},\nonumber\\
\Omega^{0'}_{\rm dm}&=&-3\Omega^0_{\rm dm}-k_c\Omega^{0'}_{\mathcal{M}}\label{eq:Om},\nonumber\\
\Omega^{0'}_{\rm b}&=&-3\Omega^0_{\rm b}\label{eq:Ob},\nonumber\\
\Omega^{0'}_{\rm r}&=&-4\Omega^0_{\rm r}\label{eq:Or},\nonumber\\
\Omega^{0'}_{\mathcal{M}} &=&Q\frac{\sqrt{6}}{\pi}{\Omega^0_{\mathcal{M}}}^{3/2}\frac{e^{-N}}{\mathcal{H}},
\end{eqnarray}
where the first one is the Friedmann equation. Observe that due to the $\pm$ sign of the 
square root of $\mathcal{M}$, we have the possibilities that $Q$ can be positive or 
negative in the evolution of $H$. 

\begin{figure*}[t!]
    \centering
    \makebox[11cm][c]{
    \includegraphics[trim = 0mm  0mm 0mm 0mm, clip, width=8.cm, height=4.5cm]{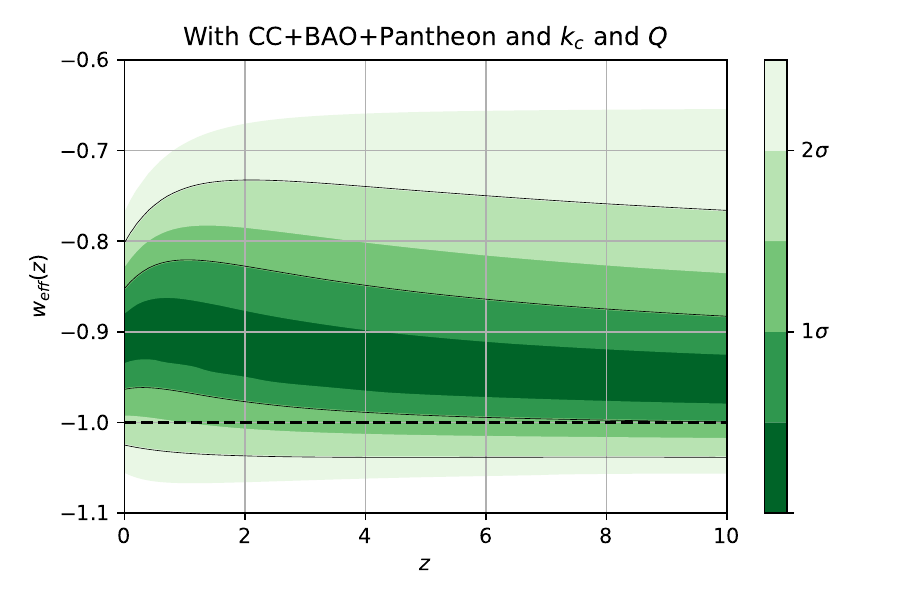}
    \includegraphics[trim = 0mm  0mm 0mm 0mm, clip, width=8.cm, height=4.5cm]{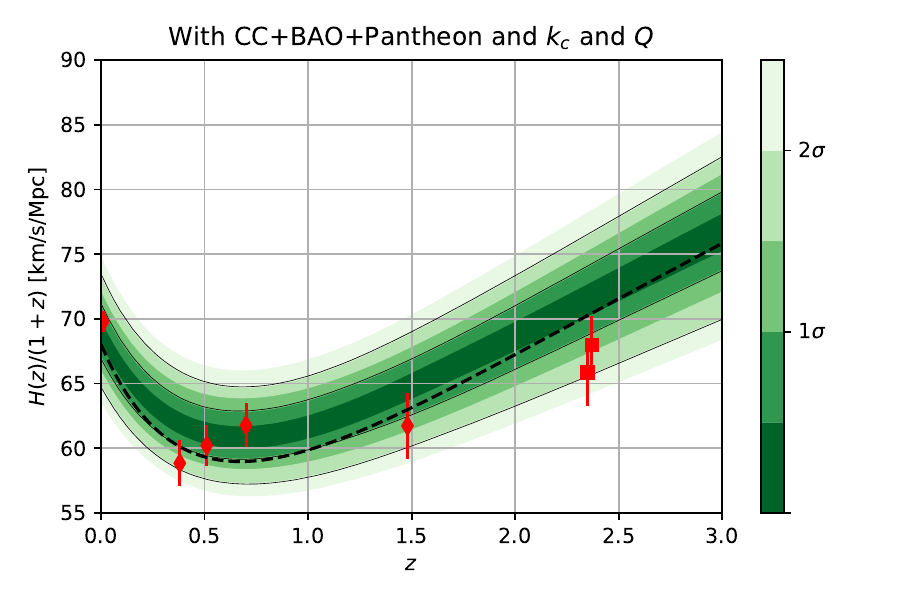}
    }
    \makebox[11cm][c]{
    \includegraphics[trim = 0mm  0mm 0mm 0mm, clip, width=8.cm, height=4.5cm]{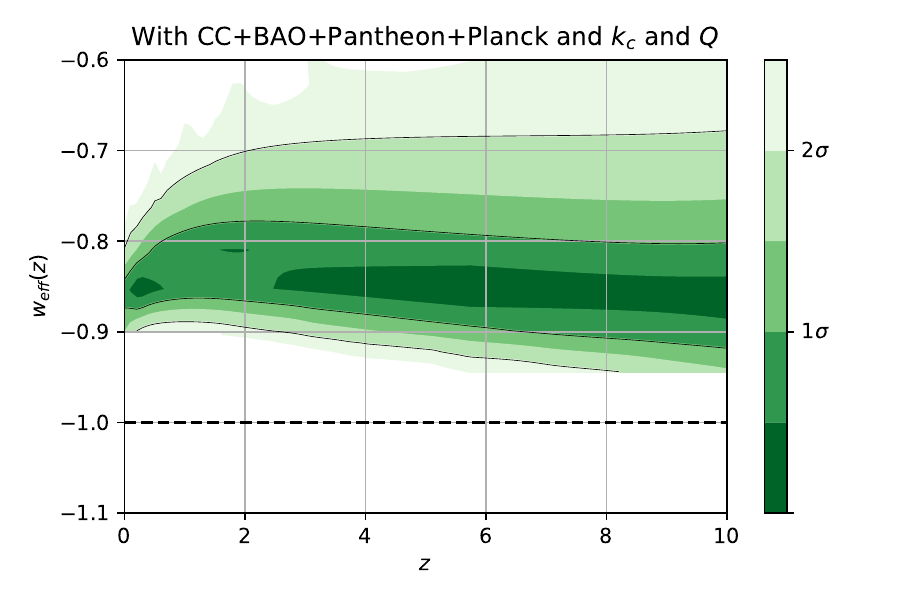}
    \includegraphics[trim = 0mm  0mm 0mm 0mm, clip, width=8.cm, height=4.5cm]{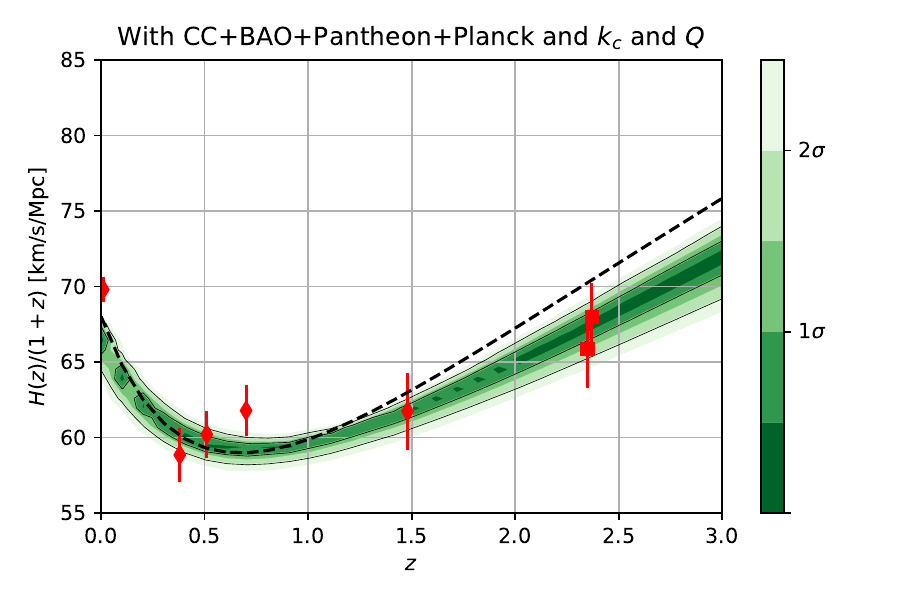}
    }
    \makebox[11cm][c]{
    \includegraphics[trim = 0mm  0mm 0mm 0mm, clip, width=8.cm, height=4.5cm]{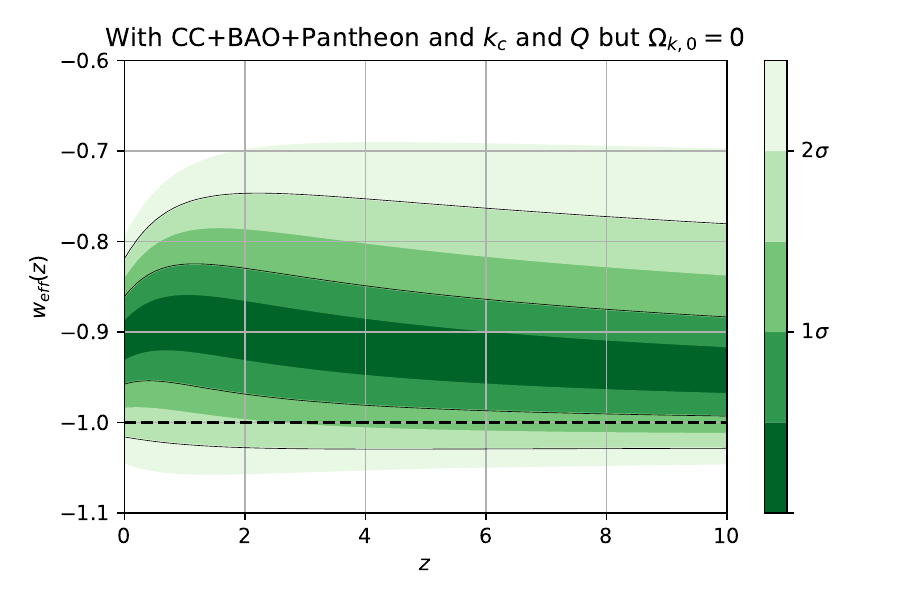}
    \includegraphics[trim = 0mm  0mm 0mm 0mm, clip, width=8.cm, height=4.5cm]{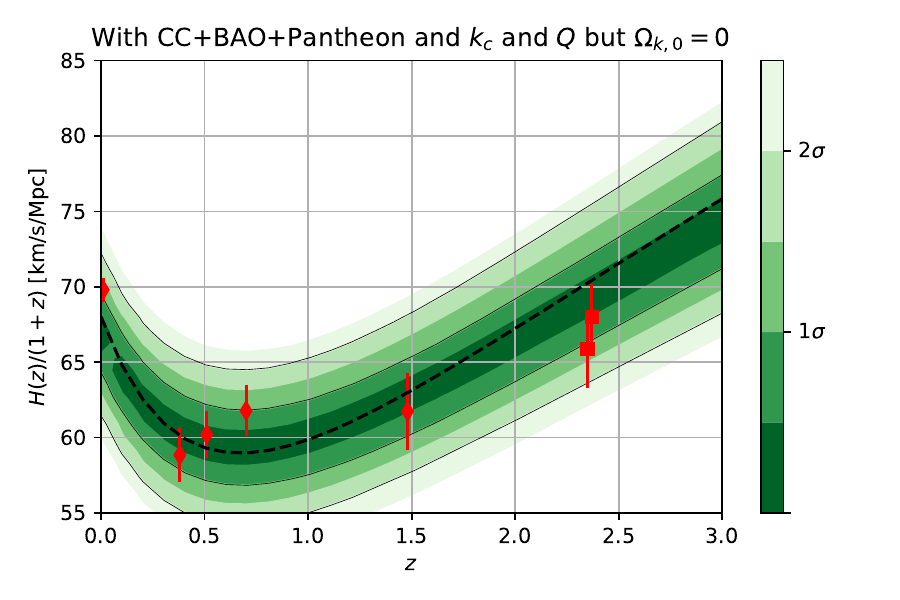}
    }
    \makebox[11cm][c]{
    \includegraphics[trim = 0mm  0mm 0mm 0mm, clip, width=8.cm, height=4.5cm]{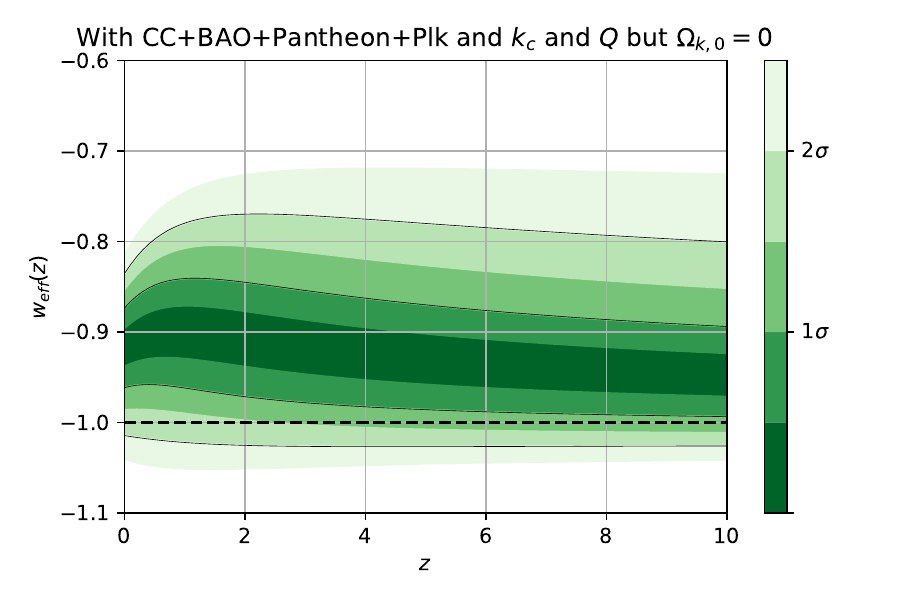}
    \includegraphics[trim = 0mm  0mm 0mm 0mm, clip, width=8.cm, height=4.5cm]{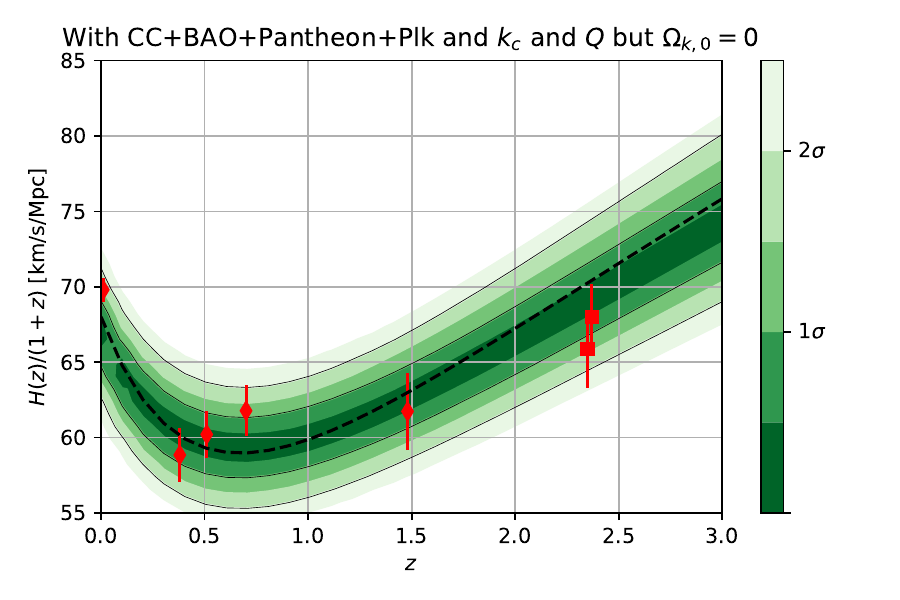}
    }
    \caption{
    Functional posterior probability of the reconstruction with flat curvature.
    The probability as normalised in each slice of constant $z$, with colour scale in confidence interval values. 
    The 68\% ($1\sigma$) and 95\% ($2\sigma$) confidence intervals are plotted as black lines. Left: the effective EoS \ref{eq:omega} and Right: the Hubble Parameter $H(z)/(1+z)$. The dashed black line corresponds to the standard $\Lambda$CDM values. 
}    %
\label{fig:cmade}
\end{figure*}

\begin{figure*}[t!]
    \centering
    \includegraphics[trim = 0mm  0mm 0mm 0mm, clip, width=12.cm, height=12.cm]{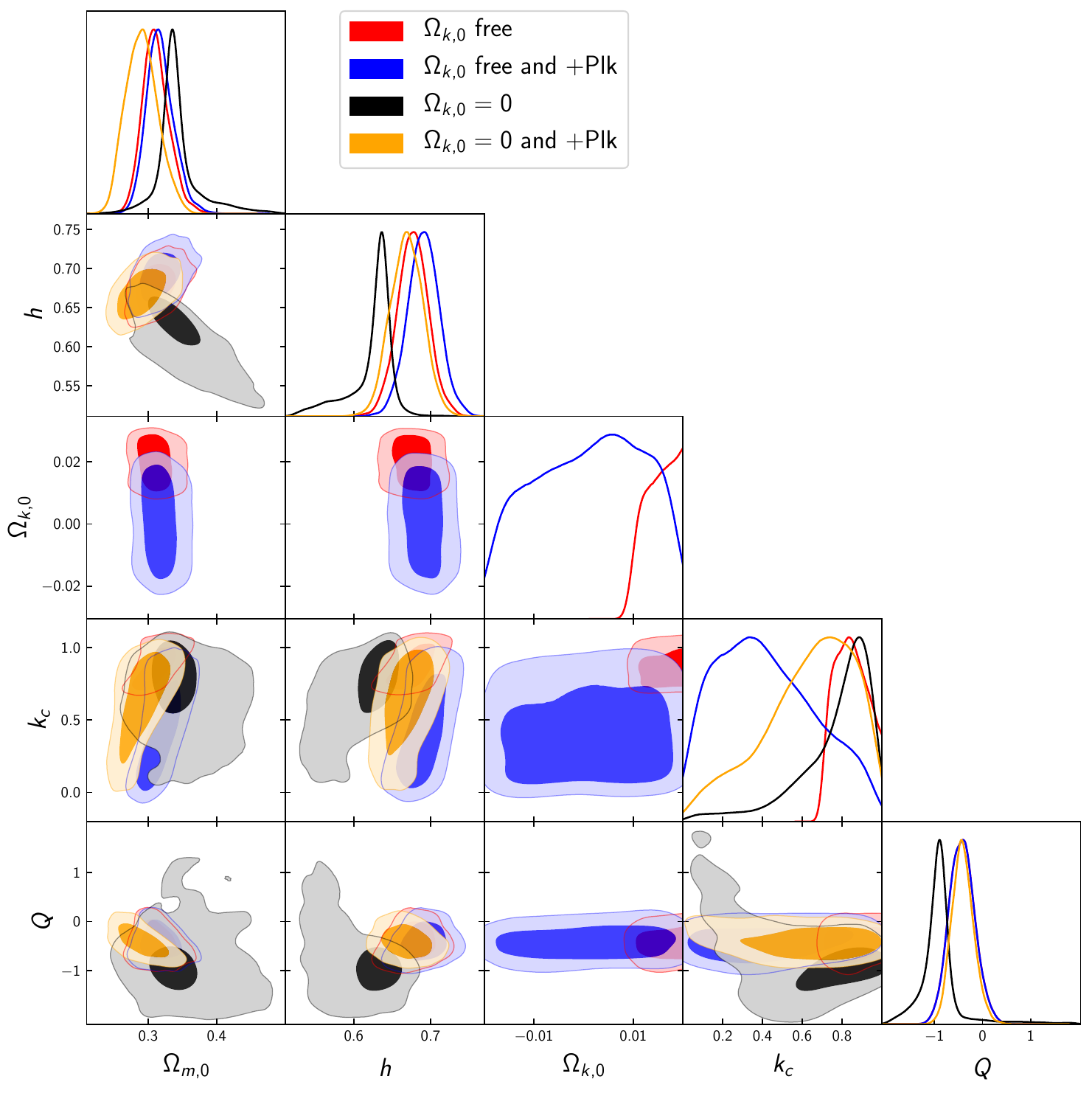}
    \caption{ 
    Triangular marginal posterior distributions for the inferred parameters; 1D posteriors are displayed
    over the diagonal and 2D below it;
    they are colour coded with the inclusion of curvature and the Planck information as shown in the labels.
}\label{fig:cmade_nocurv_plk}
\end{figure*}

\begin{figure*}
\centering
\includegraphics[trim = 0mm  0mm 0mm 0mm, clip, width=8.cm, height=5.cm]{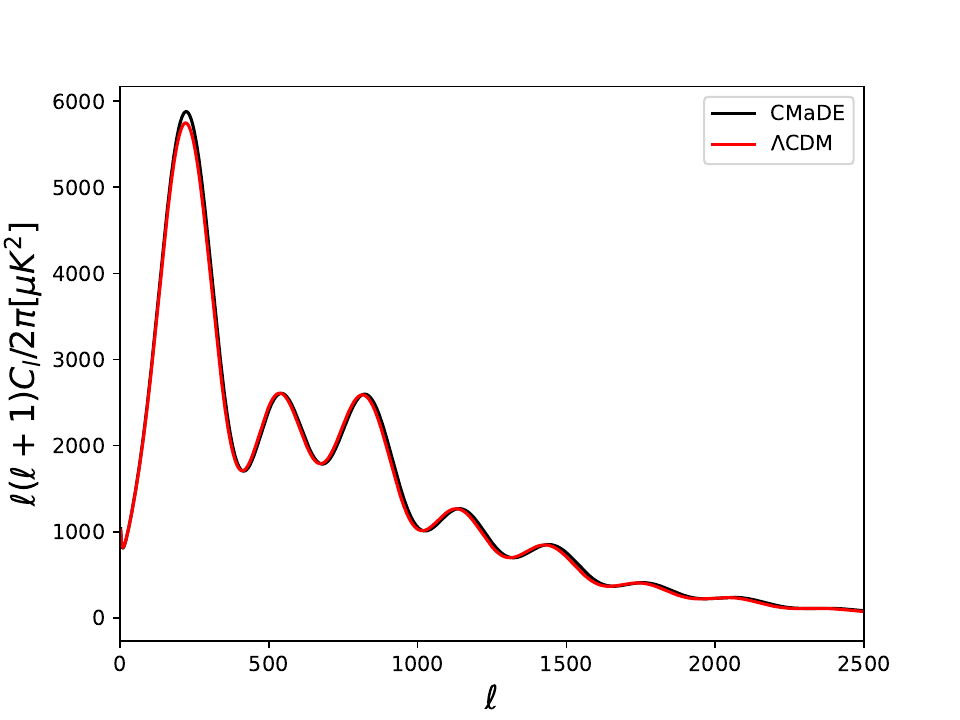}
\includegraphics[trim = 0mm  0mm 0mm 0mm, clip, width=8.cm, height=5.cm]{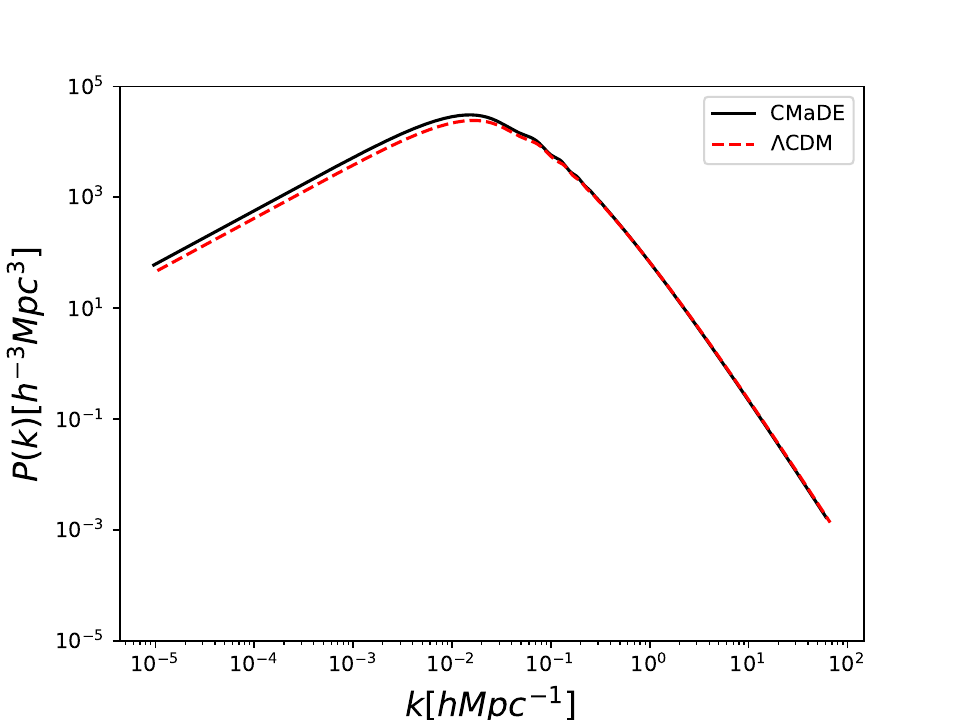}\\
\includegraphics[trim = 0mm  0mm 0mm 0mm, clip, width=8.cm, height=5.cm]{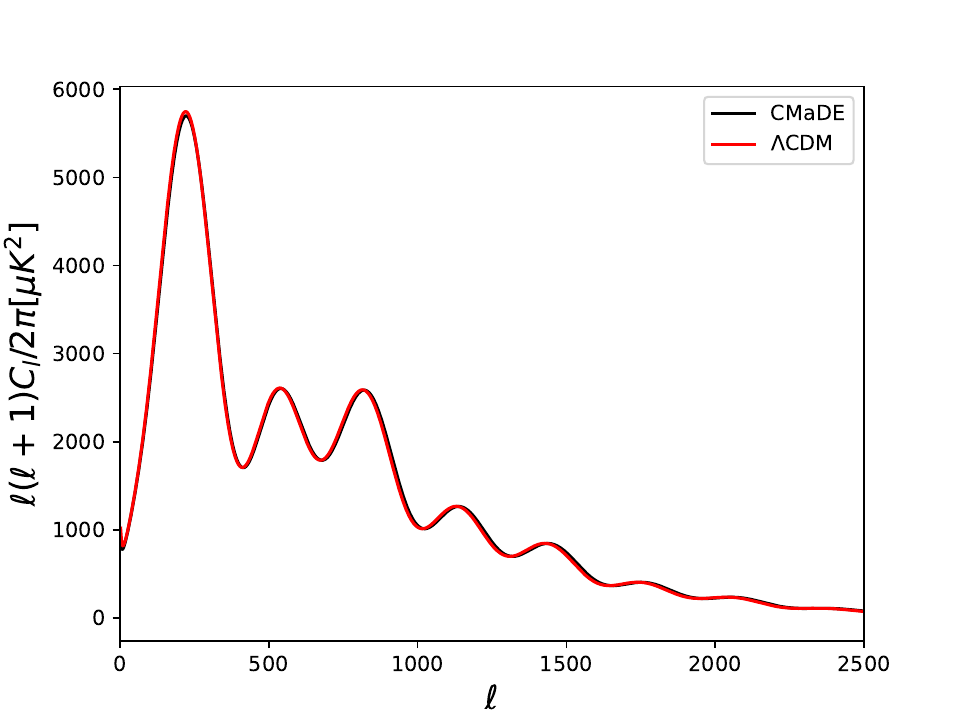}
\includegraphics[trim = 0mm  0mm 0mm 0mm, clip, width=8.cm, height=5.cm]{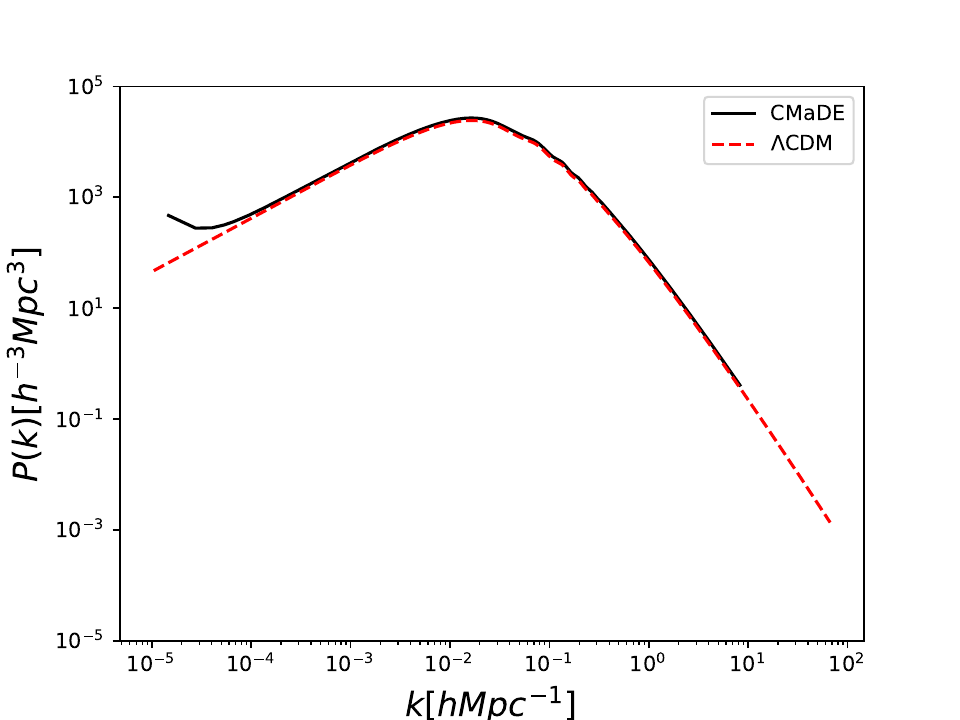}
\caption{Top: The CMB and the MPS profiles for the CMaDE model. In this figure we set $H_0=72.6$ Km/s/Mpc, 
$\Omega_{0\rm{b}}=0.044$ and $\Omega_{0\rm{k}}=0$. Bottom: The CMB and the MPS profiles for the CMaDE model. 
In this figure we set $H_0=68$ km/s/Mpc, $\Omega_{0b}=0.048$, $\Omega_{0\rm{DM}}=0.23$ and $\Omega_{0\rm{k}}=0.001$.}
\label{fig:CMB_1}
\end{figure*}

\section{Comparing with Cosmological observations}

It is straightforward to find numerical solutions of the system (\ref{eq:Sys}); 
as a test we use the initial conditions obtained from the Lambda Cold Dark Matter ($\Lambda$CDM) model, 
and implemented a python code with an Adams–Badsforth–Moulton algorithm  that integrates the system from 
$N=0$ to $N=-7$. As a proof of the concept, we set as initial conditions $\mathcal{H}_0=1$, 
$\Omega^0_{0b}=0.044$, $\Omega^0_{0\rm{dm}}=0.27$, $\Omega^0_{0\rm{r}}=9.539\times10^{-5} $ and 
$\Omega^0_{0\rm{k}}=0.08$ for both, the CMaDE and the $\Lambda$CDM  models. 
In figure \ref{fig:Claudio} we show a comparison of the Hubble parameter behaviour for the CMaDE 
and $\Lambda$CDM models. In the left panel of this figure we establish the current value of the Hubble 
constant $H_0=1$ to compare both evolutions.  
We plot the rate $1-H_{\rm{CMaDE}}/H_{\rm{LCDM}}$ in order to see the difference between both models and 
notice its difference does increase at small redshifts, within observable regions,  but converges to the $\Lambda$CDM 
at the present time and as well as at the early universe.
We think this could be an indication that the CMaDE model might ameliorate the Hubble constant tension 
by maintaining consistency with the CMB and modifying mainly the late time observables.  

Now we introduce an effective equation of state (EoS) for the CMaDE model. 
In order to do so, we define a function $w_{\rm eff}$ such that
\begin{equation}
\Omega^{0'}_{\mathcal{M}}+3(1+w_{\rm eff})\Omega^0_{\mathcal{M}}=0.
\end{equation}
We use the last equation of system (\ref{eq:Sys}) to obtain the effective equation of state
\begin{equation}\label{eq:omega}
    w_{\rm eff} =-Q\sqrt{\frac{2}{3}}\sqrt{\Omega^0_{\mathcal{M}}}\frac{e^{-N}}{\pi\mathcal{H}}-1.
\end{equation}
We plot the results in the right panel in Fig. \ref{fig:Claudio}. Note that the EoS tends to $-1$ 
in the early universe, then changes its value after recombination. We will see that the CMB values 
are very much in agreement with this behaviour. 
\\

Finally we want to compare the CMaDE model with the main cosmological observations at the background 
level, and let the perturbative study for future works. In order to do this we performed a statistical 
analysis of our model parameter space with a publicly available Bayesian inference code named 
\texttt{SimpleMC} \cite{simplemc, aubourg2015cosmological}, which includes the \texttt{dynesty} 
library~\cite{speagle2020dynesty}, a nested sampling algorithm used to compute the Bayesian evidence. 
The data sets used in this work consist of:

\begin{itemize}

    \item Baryon Acoustic Oscillations (BAO) measurements. The BAOs utilized in this study are obtained 
    from SDSS, SDSS-II, BOSS, and eBOSS. The data sets encompass the SDSS Galaxy Consensus, quasars, 
    and Lyman-$\alpha$ forests \cite{alam2017clustering, de2019baryon, ata2018clustering, blomqvist2019baryon, beutler20116df, eBOSS:2020yzd}. The sound horizon is calibrated by using Big Bang Nucleosynthesis \cite{aubourg2015cosmological}. 
    Henceforth, these measurements will be referred to as BAO.
    
    \item The complete catalog of supernovae from the Pantheon Plus SN Ia sample (referred to as SN). 
    This SN data set builds upon the original Pantheon compilation, aiming to enhance the precision and 
    inclusiveness of the supernova sample. Both the covariance matrix and the data are available in \cite{Scolnic:2021amr}.
    
    \item The Hubble Parameter, denoted as $H(z)$, is derived by compiling measurements from cosmic 
    chronometers (referred to as CC), which are old stars functioning as "standard clocks" in cosmology. 
    The CC dataset employed in this study is available in the repository \cite{hz}.

    \item We utilize data from the Planck satellite to extract information from the Cosmological Microwave 
    Background (CMB). However, our focus solely rests on the cosmological background, excluding perturbations. 
    In this context, the Planck data functions as a BAO measurement at a redshift of approximately $z\approx1100$, 
    corresponding to the last scattering epoch. This implies that we are capturing the angular scale of 
    the sound horizon at a high redshift. As elaborated in \cite{aubourg2015cosmological}, the CMB 
    information on a background level can be condensed into three parameters: $w_{\rm b}$ (physical 
    baryon density parameter), $w_{\rm cb}$ (physical matter density parameter), and $D_A(1100)/r_{\rm d}$, 
    accompanied by their associated covariance matrix.

\end{itemize}

Given that we are only focusing on background data the parameters to be used are those relevant to the 
background only. The flat priors used for these parameters are: $\Omega_{\rm m} = [0.1, 0.5]$ for the matter 
density parameter, $\Omega_{\rm b} h^2 = [0.02, 0.025]$ for the baryon density, $h = [0.4, 0.9]$ 
for the dimensionless Hubble constant $h=H_0/100\, {\rm km\,s}^{-1}{\rm Mpc}^{-1}$, the radiation is negligible 
so we will set $\Omega_{0\rm r} = 0$ and for the curvature's density parameter $\Omega_{0\rm k}=[-0.02, 0.02]$. 
For the bias parameters we choose $k_c=[0,1]$ and $Q = [-2,2]$. 
\\

The results of the parameter inference procedure can be found in Fig.~\ref{fig:cmade_nocurv_plk}: 
the marginalized posteriors for the parameters; along with Table \ref{tabla_evidencias}. 
For best-fit values (last row of the table), with their 1$\sigma$, we have: $\Omega_{\rm m} = 0.315 \pm 0.053$, 
$h=0.699 \pm 0.012$, $\Omega_{\rm k} = 0.017\pm 0.002$, $k_c = 0.83\pm0.09$ and $Q= -0.67 \pm 0.12$. 
To compare how the CMaDE model performs we assess the fitness to the data via 
$-2\Delta\ln \mathcal{L_{\rm max}}= -2\ln \mathcal{L_{\rm max,LCDM}}+2\ln \mathcal{L_{\rm max,CMaDE}}$, 
which is the difference between our model's best-fit to the data versus $\Lambda$CDM's; and the Bayes' 
Factor $B_{1,2} = E_1/E_2$, or, more specifically, its natural logarithm $\ln{B_{1,2}}$ where $E_i$ is the 
Bayesian Evidence for a model $i$. In this study we obtained $-2\Delta\ln \mathcal{L_{\rm max}}=3.65$ in favour 
of the CMaDE model, indicating a better fitness to the data used. The Bayes' factor obtained is $\ln{B_{1,2}}=2.3$, 
indicating that CMaDE is in a slight disadvantage (almost moderate evidence) against the standard model when 
explaining the observations according to the empirical Jeffrey's Scale in Table~\ref{jeffreys} following 
the convention from \cite{Trotta:2008qt}. This is not unexpected given that our model has two extra 
parameters and the Bayesian Evidence penalizes the added complexity.
\\

On the other hand, in Figure \ref{fig:cmade} we observe the functional posteriors for CMaDE's EoS 
and the Hubble Parameter $H(z)/(1+z)$. The effective EoSs present Quintessence-like behaviour, and as we go 
further into the past, CMaDE's EoS starts resembling that of LCDM as expected. The deviation of CMaDE's 
Hubble Parameter from the standard model is significant 
enough so that it fits better the BAO data at $z\approx2.35$. We consider this a positive indication 
for our model since, despite possessing a distinct theoretical foundation and dynamic behavior 
in the equation of state, our model yields characteristics that closely resemble those of the 
standard model and even explains better some observations.
\\

\begin{table}[t!]
\caption{Mean values, and standard deviations,  for the parameters used throughout the reconstructions.
For each model, the last two columns present the Bayes Factor, and the 
$-2\Delta\ln \mathcal{L_{\rm max}} \equiv -2\ln( \mathcal{L_{\rm max}}_{,\Lambda \text{CDM}} / \mathcal{L_{\rm max,}}_i$) 
for fitness comparison. Here $-2\ln \mathcal{L_{\rm max}}_{,\Lambda \text{CDM}} = 1429.71$, 
$\ln E_{\Lambda \text{CDM}}=-721.59 (0.14)$ when not including the Planck data set and 
$-2\ln \mathcal{L_{\rm max}}_{,\Lambda \text{CDM}} = 1431.4$, 
$\ln E_{\Lambda \text{CDM}}=-726.16 (0.14)$ when including it.}
\footnotesize
\scalebox{0.78}{%
\begin{tabular}{cccccc} 
\cline{1-6}\noalign{\smallskip}
\vspace{0.15cm}
Model  & $h$ &  $\Omega_{0\rm m}$  & $\Omega_{0\rm k}$ & $\ln B_{\Lambda \text{CDM},i}$  &  $-2\Delta\ln \mathcal{L_{\rm max}}$ \\
\hline
 \vspace{0.15cm}
$\Lambda$CDM            & 0.694 (0.016) & 0.311 (0.012) & --            &  0           &  0  \\
\vspace{0.15cm}
 CMaDE $\Omega_{0\rm k}=0$ & 0.646 (0.029) & 0.335 (0.058) & --            &  0.83 (0.21) & 3.21 \\
 CMaDE                  & 0.683 (0.014) & 0.311 (0.065) & 0.001 (0.011) &  0.10 (0.19) & 3.25 \\
\hline
 \vspace{0.15cm}
 With Planck data\\
  \vspace{0.15cm}
 $\Lambda$CDM           & 0.682 (0.008) & 0.302 (0.011) & --            &  0          &  0  \\
\vspace{0.15cm}
 CMaDE $\Omega_{0\rm k}=0$ & 0.685 (0.019) & 0.301 (0.051) & --            &  2.15 (0.21) & 1.35 \\
 CMaDE                  & 0.699 (0.012) & 0.315 (0.053) & 0.017 (0.002) &  2.3 (0.21) & 3.65 \\

\hline
\hline

\end{tabular}}

\label{tabla_evidencias}
\end{table}

\begin{table}[t!]
\caption{Jeffreys' scale for model selection.}
\footnotesize
\scalebox{1}{%
\begin{tabular}{cccc} 
\cline{1-4}\noalign{\smallskip}
 \vspace{0.15cm}
 $\ln{B_{12}}$ & Odds  &   Probability  &  Strength of evidence \\
\hline
\hline
\vspace{0.15cm}
$<$ 1.0 & $<$3:1 & $<$0.75 & Inconclusive  \\
\vspace{0.15cm}
1.0 &  $\sim$3:1 & 0.750 & Weak evidence  \\
\vspace{0.15cm}
2.5 & $\sim$12:1  &  0.923  & Moderate evidence  \\
\vspace{0.15cm}
5.0 & $\sim$150:1 &  0.993  & Strong evidence  \\

\hline
\hline
\end{tabular}}
\label{jeffreys}
\end{table}

For the comparison of the CMaDE model with CMB and MPS we use the CLASS code 
\cite{Lesgourgues:2011re, Matos:2021jef, Lesgourgues:2011rg, Lesgourgues:2011rh}, 
with the preferred values for $k_c=0.42$ and $Q=-0.43$ and we use an approximation similar to the one 
in \cite{Matos:2022jzf}. With this modification to the code, we generate Fig. \ref{fig:CMB_1} whose best 
fit corresponds to $H_0=68$ km/s/Mpc, $\Omega_{0\rm b}=0.048$, $\Omega_{0DM}=0.23$ and $ \Omega_{0\rm k}=0.001$. 
We see a very good coincidence with the best fit to the CMB and MPS of $\Lambda$CDM and consistency with 
the values of the previous results.

\section{Conclusions}

In this work we follow the idea of the reference \cite{Matos:2021jef} where the space-time fluctuations 
produced by the big bang are incorporated into Einstein's equations. The Einstein's equations (\ref{eq:Einstein}) 
contain an extra term $2\pi^2/{\lambda}^2$ that incorporates the energy of these fluctuations in 
space-time if ${\lambda}$ is the Compton wavelength of the graviton. In \cite{Matos:2021jef} it was 
found that these fluctuations can explain the accelerated expansion of the universe and in \cite{Matos:2022jzf} 
it was shown that these fluctuations represented in this new term in Einstein's equations are in 
agreement with the main observations of cosmology profiles of MPS and CMB. Note that the CMaDE model 
does not use an alternative theory of gravity, the only difference of the equations (\ref{eq:Einstein}) 
with Einstein's originals is the term $2\pi^2 /{\lambda}^2 $, which is a consequence of GWB fluctuations 
in space-time. We can interpret this extra term as the contribution of the GWB produced by the big 
bang to the energy of the universe.
In the present work we do not use the approximation ${T^{\mu\nu}}_{;\nu}=0$ to eliminate the small 
violation of the covariance principle, without this approximation we show that the agreement between 
the observations of CC, BAO and Pantheon is in favor with the CMaDE model, with $\Delta\chi^2=3.65$ over $\Lambda$CDM. 
In addition, we compare the CMaDE model with the CMB and MPS and we observe that the fit is again in 
agreement by using the values for the free constants of the model, reducing the so-called tension $H_0$. 
The final conclusion is that the accelerated expansion of the universe can be explained taking into account 
the GWB energy of space-time, without cosmological constant or modifications of Einstein's equations.

\section*{Acknowledgements}
We thank Luis Osvaldo Tellez for his advice and help with the python codes. This work was partially 
supported by CONACyT M\'exico under grants  A1-S-8742, 304001, 376127, 240512,  FORDECYT-PRONACES grant 
No. 490769 and I0101/131/07 C-234/07 of the Instituto Avanzado de Cosmolog\'ia (IAC) collaboration (http://www.iac.edu.mx/).
JAV acknowledges the support provided by FOSEC SEP-CONACYT Investigaci\'on B\'asica A1-S-21925, 
UNAM-DGAPA-PAPIIT IN117723 and FORDECYT-PRONACES-CONACYT/304001/2019.

\bibliographystyle{ieeetr}
\bibliography{referencias}{}

\end{document}